\documentclass[aps,prl,twocolumn,superscriptaddress,fleqn,a4paper]{revtex4-1}
\usepackage[ngerman,english]{babel}
\usepackage[utf8]{inputenc}   
\usepackage{xspace}
\usepackage{amsmath,tensor}
\usepackage{array,color}
\usepackage{relsize,exscale}  
\usepackage{textcase}	
\usepackage{soul,color}		
\usepackage{calc} 
\usepackage{makeidx,showidx}
\usepackage{graphicx}
\usepackage{wrapfig}
\usepackage{float}
\usepackage{textcomp}
\usepackage{siunitx}
\usepackage[version=3]{mhchem}
\usepackage{enumitem}
\usepackage{fancyref}
\usepackage[hidelinks=true]{hyperref}
\sethlcolor{green}
\setulcolor{red}
\usepackage{lineno}			

\newcommand{\erww} [1] {\ensuremath{\langle {#1} \rangle}}

\newcommand{\lsco} {{La$_{2-x}$Sr$_x$CuO$_4$}\@\xspace}

\newcommand{\hgryb} {{HgBa$_{2}$CuO$_{4+\delta}$}\@\xspace}

\newcommand{\ybcoF} {$\ce{YBa2Cu3O_{7}}$\@\xspace}
\newcommand{\ybco} {$\ce{YBa2Cu3O_{6+y}}$\@\xspace}

\newcommand{\ybcoE} {$\ce{YBa2Cu4O8}$\@\xspace}

\newcommand{\tc} {\ensuremath{T_{\rm c}}\@\xspace}
\newcommand{\cperp}{\ensuremath{c \bot B_0}\@\xspace}
\newcommand{\cpara}{\ensuremath{{c\parallel\xspace B_0}}\@\xspace}

\newcommand{\beq} {\begin{equation}}
\newcommand{\eeq} {\end{equation}}

\newcounter{exex}[section]

\makeatletter\newcommand\listofexamples{\section*{List of Examples}\@starttoc{xmp}}
	\newcommand\l@example[2]{\par\noindent#1~\textit{#2}\par}
\makeatother






\makeatletter
\renewcommand\subsection{\@startsection 
{subsection}{3}{0mm}
{-\baselineskip}
{0.5\baselineskip}
{\centering \textbf }}
\makeatother

\makeatletter
\renewcommand\subsubsection{\@startsection 
{subsubsection}{3}{0mm}
{-\baselineskip}
{0.5\baselineskip}
{\centering  }}
\makeatother

\makeindex

\begin{document}
\title{Unconventional $^{17}$O and $^{63}$Cu NMR shift components in cuprate superconductors}
\author{Danica Pavi\'cevi\'c}
\author{Marija Avramovska}
\author{Jürgen Haase}
\affiliation{University of Leipzig, Felix Bloch Institute for Solid State Physics, Linn\'estr. 5, 04103 Leipzig, Germany}

\begin{abstract}
Nuclear magnetic resonance (NMR) is a fundamental bulk probe that provides key information about the electronic properties of materials. Very recently, the analysis of all available planar copper shift as well as relaxation data proved that while the shifts cannot be understood in terms of a single temperature dependent spin component, relaxation can be explained with one dominating Fermi liquid-like component, without enhanced electronic spin fluctuations. For the shifts, a doping dependent isotropic term, as well as doping independent anisotropic term became obvious. Here we focus on planar $^{17}$O NMR shifts and quadrupole splittings. Surprisingly, we find that they demand, independently, a similar two-component scenario and confirm most of the previous conclusions concerning the properties of the spin components, in particular that a negative spin polarization survives in the superconducting state. This should have consequences for the pairing scenario.
\end{abstract}

\maketitle
\vspace{0.5cm}
{\centering \today}
\labelformat{paragraph}{#1}

\subsection{Introduction}
Soon after the discovery of cuprate superconductivity \cite{Bednorz1986} nuclear magnetic resonance (NMR), which was the first probe to prove BCS theory \cite{Hebel1957,Bardeen1957}, uncovered essential nuclear shift and relaxation data that theory tried to turn into a physical picture of these materials (for reviews of cuprate NMR see \cite{Slichter2007,Walstedt2008}). 

Among the important early NMR findings was spin singlet pairing that was deduced from the vanishing of some low-temperature shifts and relaxation. A so-called spin gap was discovered, as well, since for the underdoped systems the shifts were temperature dependent already far above the superconducting transition temperature \tc, different from what one expected from the uniform response ($\chi_0$) of a Fermi liquid (Pauli susceptibility). 
\begin{figure}[t]
\centering
\includegraphics[width=0.45\textwidth ]{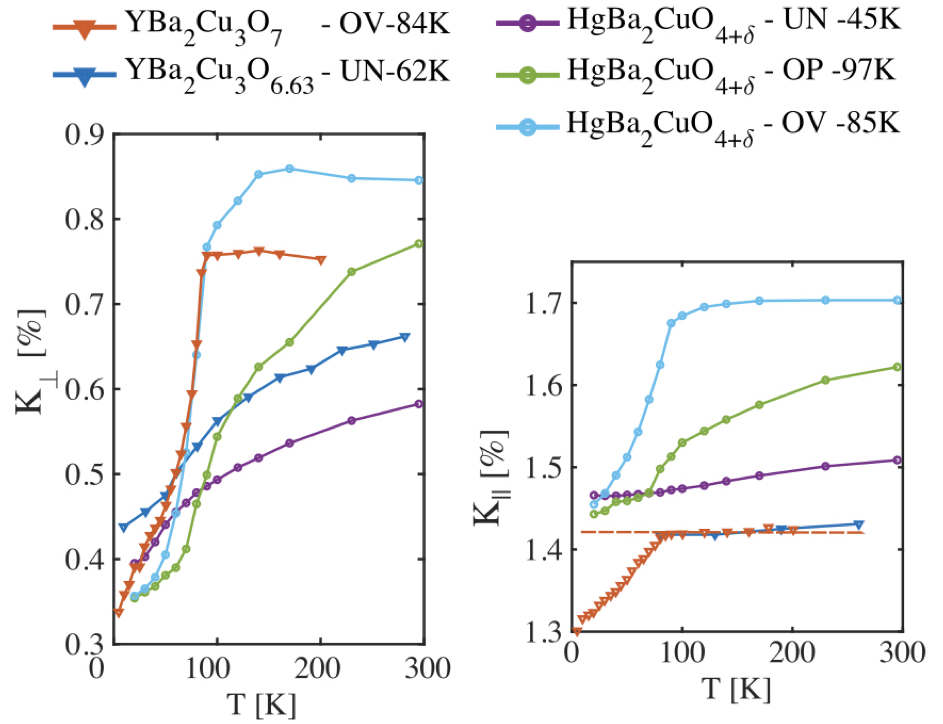}
\caption{Typical planar $^{63}$Cu shift data in the curates for 5 different samples \cite{Rybicki2015,Takigawa1989,Takigawa1991} and two directions of the external field, parallel (\cpara) and perpendicular (\cperp) to the crystal $c$-axis, $K_{\parallel,\perp}$ (total shifts are shown). \emph{Left panel:} Optimally (OP) or overdoped (OV) systems have a rather temperature independent $K_\perp$ that rapidly decreases below \tc to reach some similar low temperature values. As the doping level is lowered the shifts begin to decrease already at increasingly higher temperatures above \tc. \emph{Right panel:} $K_\parallel$ for the same materials; note that for the \ybco family of materials this shift was considered largely temperature independent after a diamagnetic correction \cite{Barrett1990} (dashed line) and represent the orbital shift value for \cpara (for \ybcoE it is temperature independent without the correction \cite{Bankay1994}); some materials show rather large temperature dependent $K_\parallel$, pointing to rather different hyperfine coefficients in the old scenario.}.\label{fig:fig1}
\end{figure}
It was also concluded that NMR favors a single-fluid description \cite{Takigawa1991,Bankay1994}, as shifts measured at planar Cu (for one orientation) and O seemed to follow the overall temperature dependence of the uniform magnetic susceptibility. This fluid appeared to violate the Korringa ratio between nuclear relaxation and shift, a hallmark of Fermi liquids, which was explained with enhanced electronic spin fluctuations that increase nuclear relaxation of planar Cu compared to what one would expect from its spin shift \cite{Walstedt1988,Mila1989b}. 

Unfortunately, most of the conclusions had to be drawn from a small number of materials, and the \ybco family was rather instrumental. The reason for this is not only that it can have a high transition temperature, but also that other cuprates have excessive NMR linewidths that make measurements tedious and question the quality of the crystals. 

\begin{figure}
\centering
\includegraphics[width=0.45\textwidth ]{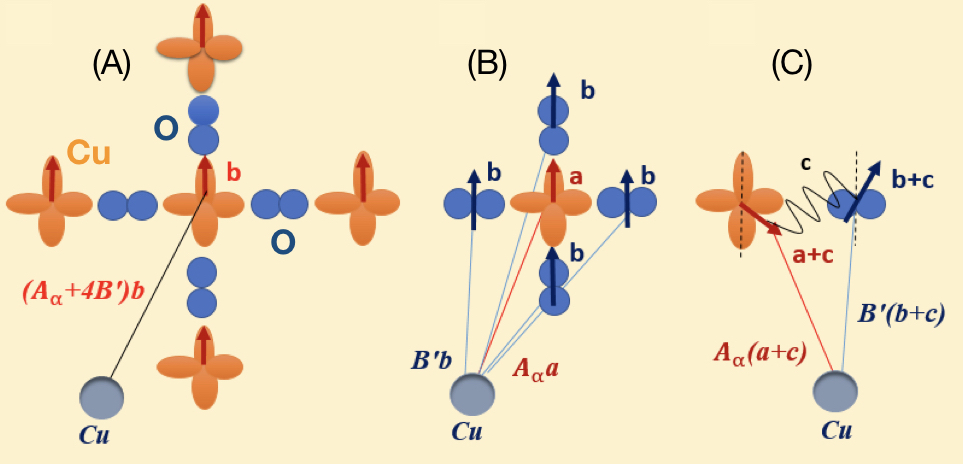}
\caption{Schematic hyperfine coupling of a Cu nucleus to the electronic spin in the cuprates. (A) The single spin component $b$ in a single band view and the old hyperfine scenario $(A_\alpha+4B')b$. (B) Two different, uncoupled spin components $a$ and $b$ couple to the nucleus through two different hyperfine coefficients $A_\alpha$ and $B'$. (C) A coupling $c$ between $a$ and $b$ can reduce the uniform response.}\label{fig:fig2}
\end{figure}

A particular early conundrum was a missing, negative spin shift for Cu NMR from an expected spin in the partially filled $3d(x^2-y^2)$ orbital when the external field is parallel to the crystal $c$-axis (\cpara), i.e. for $K_\parallel$. In fact, for this orientation it was concluded that there is no spin shift at all. And since $K_\perp$ did show a temperature dependence, cf.~Fig.~\ref{fig:fig1}, it was assumed that there is an accidental cancellation from \emph{two} hyperfine coefficients for the Cu shifts, 
\beq\label{eq:single}
K_{\parallel,\bot}(T) = (A_{\parallel,\bot}+4B')\cdot \chi_0(T),
\eeq
i.e., $A_{\parallel}+4B' \approx 0$, for spin from the onsite atom ($A_{\parallel,\perp}$) and the four neighboring Cu atoms ($B'$) in a single band scenario, cf.~Fig.~\ref{fig:fig2} (A). One should note that there were slight ambiguities with respect to the true $K_\parallel(T\rightarrow 0)$ due to the size of the diamagnetic correction considered \cite{Barrett1990}, cf. Fig.~\ref{fig:fig1}.
With the appearance of Cu shift data that showed a significant temperature dependence for \cpara, cf.~Fig.~\ref{fig:fig1}, the old hyperfine scenario was not revisited. Over the last decade, with a number of special experiments these assumptions were questioned, and a two-component scenario was proposed \cite{Haase2009b,Meissner2011,Haase2012,Rybicki2015}. In such a scenario, two different spin components couple to each nucleus with two different hyperfine coefficients and have different temperature dependences, cf.~Fig.~\ref{fig:fig2}(B). In general, a coupling between the two spin components is to be expected and this leads to another coupling term, cf.~Fig.~\ref{fig:fig2}(C). 

Finally, by just reviewing the literature body of Cu shifts (these are most reliable even in the mixed state since they are large), a rather different Cu shift phenomenology appears from simple plots \cite{Haase2017}, and the failure of a single spin component description becomes obvious. 

For example, an \emph{isotropic} Cu shift term becomes visible as a function of doping, which says that a large amount of $K_\parallel$ should be spin shift, as well, and not orbital shift. One also notices that $K_\parallel$ is always greater than $K_\perp$, so that one is lead to assume a negative spin component that, together with the very anisotropic hyperfine constant from the $3d(x^2-y^2)$ orbital ($|A_\perp| \ll |A_\parallel|$ and $A_\parallel < 0$)  produces an additional positive shift only for $K_\parallel$. The particular temperature dependent shift variations are more complicated, but  seem to obey simple rules, as well \cite{Haase2017}.

Another very recent discovery concerns the Cu nuclear relaxation. By inspecting all literature data \cite{Avramovska2019,Jurkutat2019} it becomes apparent that there is no reason to invoke special electronic spin fluctuations that are very different from those of a Fermi liquid. Rather, it was argued by APH \cite{Avramovska2019} that the spin shifts must be suppressed for most cuprates (and not the relaxation enhanced), causing the Korringa relation to fail. This can be understood with the two spin susceptibilities, as well, cf. Fig.~\ref{fig:fig2}. In particular, APH argued that the two spin components couple through $A_{\parallel,\perp}$, the Cu 3$d(x^2-y^2)$ term, and an isotropic transferred coupling $B$, i.e.,
\beq\label{eq:two1}
K_{\bot,\parallel} (T)= A_{\bot,\parallel}\cdot  \chi_1(T) + B \cdot \chi_2(T),
\eeq
we prefer to use $B = 4B'$ compared to the old picture. Then, the two susceptibilities lead to two spin polarizations in a magnetic field: $(a+c) \equiv \chi_1$ and $(b+c) \equiv \chi_2$, where $c$ represents a coupling term that can always be present \cite{Haase2009b}. One then has,
\beq\label{eq:two2}
K_{\bot,\parallel} (T)= A_{\bot,\parallel}\cdot  (a+c) + B \cdot (b+c),
\eeq
and a negative $c$ may cause an antiferromagnetic alignment of two uncoupled spins $a$ and $b$, and lead to a reduced spin response. This is illustrated in Fig.~\ref{fig:fig2}(C).

Here we would like to focus on $^{17}$O NMR data. While sparse due to the necessary isotope exchange, we find with the new understanding of the charge distribution in the \ce{CuO2} plane that the rather unusual $^{17}$O NMR shifts and lineshapes that were discussed, but not understood, nearly 20 years ago, are in direct support of a the basic model proposed by APH \cite{Avramovska2019}, and in particular a doping dependent isotropic Cu shift that relates to the charge. That is, a proper understanding of shift and splitting demands a two-component spin response, even a negative $^{17}$O shift that is a direct consequence of the negative spin component proposed by APH is found. 

\subsection{NMR shift contributions}
Before we discuss the shifts in more detail, we need to point to the various NMR shift terms. For a particular direction ($\alpha$) of the magnetic field with respect to the crystal axes we can simply write,
\beq\label{eq:shift1}
\hat{K}_\alpha = K_{\rm core}+K_{L,\alpha}+K_\alpha(T)+K_{\rm dia, \alpha}(T).
\eeq
The first two terms are orbital shifts, from the core electrons ($K_{\rm core}$) and from bonding orbitals ($K_{L,\alpha}$, also called van Vleck term). The third term is the spin shift ($K_\alpha$), due to a rather isotropic electronic spin, which acquires its orientational dependence from hyperfine coefficients. A partial diamagnetic response from the superfluid in the mixed state is described by ($K_{\rm dia, \alpha}$). 
While we do not know $K_{\rm dia,\alpha}$ precisely, it can safely be neglected for the large Cu NMR and even the planar O in typical magnetic fields of several Tesla where the penetration depth is large compared to the distance between fluxoids. Experimentally, we know this from the much smaller NMR shifts and broadenings of Y, Hg, or apical O that must experience very similar $K_{\rm dia,\alpha}$, i.e., their shifts and broadenings represent an upper limit for the effect. 

Then, if one chooses a suitable reference compound with small $K_{L,\alpha}$ and $K_\alpha$, the measured total shift (relative to the reference) is basically given by,
\beq\label{eq:shift02}
\hat{K}_\alpha(T)= K_{\rm L,\alpha}+K_\alpha(T),
\eeq 
i.e. an orbital and a spin shift component (for related problems with definitions in the literature cf.~\cite{Haase2017}). 

In addition to the shifts there is the electric quadrupole interaction that splits the $^{63,65}$Cu and $^{17}$O NMR resonances ($I=3/2, I=5/2$) into $2I+1$ lines. However, this splitting can be measured so that the quadrupole effects can be largely removed from the shifts (see below). In fact, the quadrupole splittings in the cuprates are determined quantitatively by the planar Cu and O hole densities \cite{Haase2004}.
 
As mentioned in the introduction, for \ybco one finds for Cu that ${^{63}\hat{K}}_\parallel$ is nearly temperature independent while ${^{63}\hat{K}}_\perp$ is not. So it was concluded, early on,
\beq \label{eq:shift03}
\begin{split}
&{^{63}\hat{K}}_\parallel = {^{63}{K}}_{\rm L,\parallel}\\
&{^{63}\hat{K}}_\bot(T) = {^{63}K}_{\rm L,\bot}+ {^{63}K}_{\bot}(T).
\end{split}
\eeq
However, it was pointed out \cite{Pennington1989} that the thus defined orbital shift anisotropy of ${^{63}{K}}_{\rm L,\parallel}/{^{63}{K}}_{\rm L,\bot}\sim 4$ fits a single Cu ion value, but should be far too large for a hybridization with planar O (a value of 2.4 is perhaps more appropriate \cite{Renold2003}) .

\begin{figure}
\centering
\includegraphics[width=0.25\textwidth ]{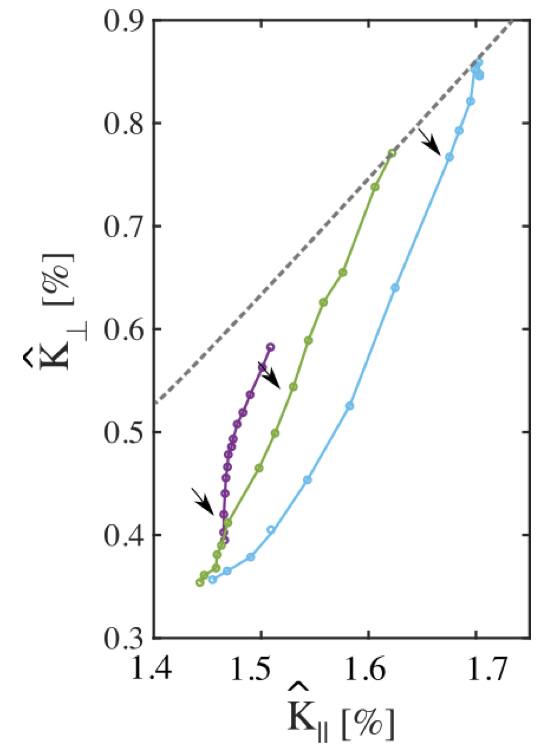}
\caption{Typical element of a $\hat{K}_\perp$ vs. $\hat{K}_\parallel$ plot with temperature as an implicit parameter, here for the \hgryb family of materials (adopted from \cite{Rybicki2015}). All shift points are located below a nearly isotropic shift line (dashed line with slope ${\sim1}$). For higher doping the temperature independent high-temperatures shifts are all located on the isotropic shift line, cf. also Fig.~\ref{fig:fig1}, but depart from it at a temperature (NMR pseudogap temeprature) that can be larger than \tc (indicated by the arrows).  Some underdoped systems may not reach up to the isotropic shift line (limited data).}\label{fig:fig3}
\end{figure}

With a rather isotropic uniform spin response and single spin component $K_\alpha = H_\alpha \chi_0$ it was assumed,
\begin{gather}
H_\parallel = A_\parallel + 4 B' \approx 0 \label{eq:hf1}\\
H_\perp = A_\perp + 4 B' > 0,  \label{eq:hf2}
\end{gather}
where $A_{\parallel} , A_\bot$ are the core polarization hyperfine coefficients for a spin in the 3d$(x^2-y^2)$ orbital, which have the very reliable property \cite{Pennington1989}, 
\beq\label{eq:A}
A_\parallel = -| A_\parallel|$ and $|A_\parallel | \gtrsim 6 A_\bot. 
\eeq
When cuprates with a substantial temperature dependent shift for \cpara appeared, i.e., ${^{63}\hat{K}}_\parallel (T) =  {^{63}K}_{\rm L,\parallel}+ {^{63}K}_\parallel (T)$, the accidental cancellation did not hold. And, if one plots their total shifts ${^{63}\hat{K}}_\bot$ vs. ${^{63}\hat{K}}_\parallel$ (without subtracting any orbital shifts components) one sees that ${^{63}\hat{K}}_\bot (T)$ is \emph{not} a linear function of ${^{63}\hat{K}}_\parallel (T)$ \cite{Rybicki2015}. This is shown for the \hgryb family of materials in Fig.~\ref{fig:fig3} \cite{Rybicki2015}. Note that a high-temperature offset between parallel lines is found (except for the lowest doping), but the offset disappears at low temperatures. In a single component view this cannot be understood since a single temperature dependent spin component must lead to proportional shift changes, i.e., 
\beq\label{eq:single}
\Delta K_\parallel/\Delta K_\perp = H_\parallel/H_\perp = const.
\eeq
which is clearly not the case.

\subsection{Unconventional planar Cu spin shift}
\begin{figure}
\centering
\includegraphics[width=0.4\textwidth ]{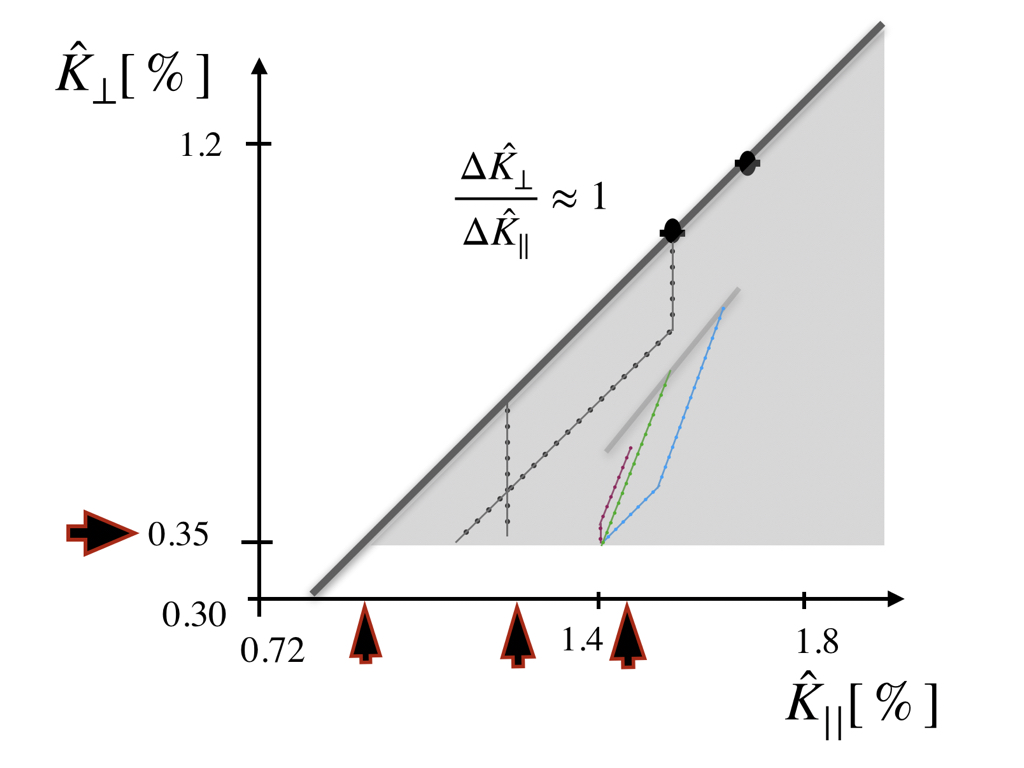}
\caption{Schematic representation of total planar $^{63}\hat{K}_{\bot,\parallel}$ Cu shifts, plotted against each other as function of temperature that is an implicit parameter, as well as doping.  Main observations are: (a) All shifts are found in the shaded lower right triangle. (b) The presence of isotropic shift lines, i.e. isotropic change of both shifts with respect to doping at high temperatures. (c) The presence of various $\Delta_T {^{63}\hat{K}}_\perp (T)/ \Delta_T {^{63}\hat{K}}_\parallel (T)$ slopes; for example the steep slope, as seen for \ybco, a slope of 2.5 for \hgryb, and a slope 1 at low T for highly overdoped Tl-based systems. (d) All low temperature ($T \rightarrow 0$) $^{63}\hat{K}_\perp$ reach a similar value of ${^{63}K_{L,\perp}} \approx 0.35 \%$, the perpendicular orbital shift value (arrow to the left). However,  $^{63}\hat{K}_\parallel (T \rightarrow 0)$ shows a larger distribution (indicated by arrows on the abscissa). The intersection of an isotropic shift line with  $^{63}\hat{K}_\perp (T\rightarrow 0)$ could define ${^{63}K_{L,\parallel}}$ not far from the predicted orbital shift anisotropy of 2.4 \cite{Renold2003}.}\label{fig:fig4}
\end{figure}

By plotting all available literature Cu shifts similar to what is shown for \hgryb  in Fig.~\ref{fig:fig3}, various observations were made, which become obvious in their figure 7 \cite{Haase2017}; a sketch in Fig.~\ref{fig:fig4}  helps summarizing them.

Proportional shift changes (straight lines in Fig.~\ref{fig:fig4}) as required by a single component view \eqref{eq:single} are restricted to small intervals of doping or temperature so that a two-component picture has to be adopted. Obviously, all experimental points as function of doping or temperature can be found below \emph{isotropic shift lines} \cite{Haase2017}, i.e., in the lower right triangle of the ${^{63}\hat{K}}_\perp$ vs. ${^{63}\hat{K}}_\parallel$ plot, cf.~Fig.~\ref{fig:fig4}. This line with slope close to 1 is formed by the high-temperature, doping dependent shifts for both orientations of the field, i.e.,
\beq\label{eq:iso1}
\frac{{^{63}\hat{K}}_{\perp}(x_2)-{^{63}\hat{K}}_{\perp}(x_1)}{{^{63}\hat{K}}_{\parallel}(x_2)-{^{63}\hat{K}}_{\parallel}(x_1)}\approx 1,
\eeq
where $x_i$ refers to particular doping levels. While unlikely on general grounds, even if one tries to explain this finding with an orbital shift variation, the slope of 1 cannot be understood. Firstly, a factor of $\sim4$ follows for an isolated \ce{Cu^{2+}} ion \cite{Pennington1989}, and a value of 2.4 is more realistic for a hybridization with oxygen \cite{Renold2003}. Secondly, ${^{63}\hat{K}}_\perp(T\rightarrow 0)$ is rather doping and family independent, pointing to a largely doping and family independent  ${^{63}{K}}_{L,\bot}$, while ${^{63}{K}}_{L,\parallel}$ differs significantly between different materials (arrows in Fig.~\ref{fig:fig4}).

On the other hand, we do expect an isotropic hyperfine coefficient $B$ to be present on general grounds, and it would even demand the observed dependence. This means that ${^{63}\hat{K}}_\parallel$ is partly spin shift, not orbital shift, which is an important conclusion. APH \cite{Avramovska2019} showed that these findings can be understood in a two-component scenario \eqref{eq:two2}, by suitable variations of $a, b, \text{and } c$, and that a negative spin component $a+c$ is present (inverted with respect to the external field, e.g., from the coupling with a dominating positive spin component $b$, cf.~Fig.~\ref{fig:fig2}(C)). This leads to the suppression of the spin shift compared to what one would expect from a Fermi liquid-like relaxation (note that the relaxation was found to be rather ubiquitous for all cuprates, except \lsco that has additional relaxation \cite{Avramovska2019,Jurkutat2019}). Therefore, it is desirable to have independent evidence for the new scenario.

\subsection{Planar charge distribution and $^{63}$Cu and $^{17}$O quadrupole splittings}
 
Today we know that the macroscopic chemical doping $x$ can be calculated from the average local hole densities of the Cu 3d$(x^2-y^2)$ and the oxygen 2p$_\sigma$ bonding orbitals, $n_d$ and $n_p$, respectively, which are measured with NMR \cite{Haase2004,Jurkutat2014}. As expected, 
\beq\label{eq:x}
x = \erww{n_d} + 2\erww{n_p} - 1.
\eeq
In parent materials $x=0$, but $n_p$ and $n_d$ differ between families. Upon doping, holes ($x > 0$) or electrons ($x < 0$) enter the \ce{CuO2} plane in a family specific way \cite{Jurkutat2014}, and the sum of the doped hole content $x = \Delta \erww{n_d} + 2\Delta \erww{n_p}$ leads to the well-known dome-like dependence of \tc on $x$ with the maximum occurring near $x \approx 15\%$, i.e. also if measured with NMR, and this value is independent on family \cite{Jurkutat2014}. However, the maxima of the domes for different cuprate families are proportional to the corresponding parent's $\erww{n_p}$ and thus $\erww{1- n_d}$, i.e. the relocation of charge from planar O to Cu raises the maximum achievable \tc \cite{Jurkutat2014,Rybicki2016}. In fact, a number of cuprate properties appear to be determined by $n_p$ or $n_d$, rather than $x$ \cite{Jurkutat2019}. This casts doubt on the efforts to discuss the cuprate properties in the old $T_c(x)$ phase diagram, only \cite{Rybicki2016}.

The hole contents $n_d$ and $n_p$ follow from the NMR quadrupole splittings \cite{Haase2004,Jurkutat2014},
\begin{gather}
{^{63}\nu}_{\rm Q,c} \approx \SI{94.3}{MHz}\cdot n_d - \SI{5.7}{MHz}\cdot (8-4 n_p)\label{eq:cufreq}\\ 
{^{17}\nu}_{\rm Q,c} \approx \SI{1.27}{MHz}\cdot n_{ p}+\SI{0.2}{MHz} . \label{eq:oxfreq}
\end{gather}
Note that spatial variations of hole contents lead to quadrupolar linewidths, i.e. distributions in ${^{n}\nu}_{\rm Q,c}$,  which for planar O depends only on $n_p$, according to \eqref{eq:oxfreq}. 

Typical $^{63}$Cu satellite linewidths are 10\% for \lsco, 1\% for \ybco, and 12\% for \hgryb \cite{Rybicki2009}, however, since the linewidths seem to vary with crystal preparation, crystal quality was often associated with it. Interstitial or mixed valence doping are likely to cause inhomogeneities for local probes, and indeed, the stoichiometric compounds \ybcoE and \ybcoF show the smallest linewidths. Furthermore, co-doping other ions increases the linewidths, as well. For electron doped systems, the linewidths can even exceed the small average splitting \cite{Jurkutat2013}. In view of \eqref{eq:cufreq} it is clear that even small variations of $n_d$ lead to large Cu linewdiths (a 2.8\% hole variation on Cu corresponds to \SI{3}{MHz} of linewidths observed for \hgryb).

Typical $^{17}$O quadrupole satellite linewidths range from 
about 1\% to 3\% \cite{Jurkutat2019}. Interestingly, the materials with the smallest Cu linewidths, \ybcoE and \ybcoF, show split oxygen satellites. Only very recently, it was proven experimentally \cite{Reichardt2018} that the $^{17}$O NMR double peak pattern in \ybco is not due to the orthorhombicity, as widely assumed, but indeed the result of an intra unit cell charge density variation of planar oxygen charge that was proposed already 2003 \cite{Haase2003}. It can also be responsible for the linewidths of the planar Cu through \eqref{eq:cufreq}. This charge density variation is about $\pm$1\% of the total oxygen charge, and can be larger in some systems, and there is good reason to believe that this kind of intra cell charge variation is ubiquitous to the cuprates \cite{Jurkutat2019}.

In order to be able to discuss the local charges and their variation and not to confuse it with the average doping $x$, we introduce a new variable, 
\beq\label{eq:zeta}
\zeta=n_d+2n_p-1,
\eeq
that is the local hole content for a particular \ce{CuO2} unit of the material. Then,
\beq\label{eq:zetax}
x = \erww{\zeta} = \left<n_d\right> + 2\left< n_p\right>-1.
\eeq
Note that the intra-cell charge variation is not captured by $\zeta$.


\subsection{Unconventional spin shifts and planar charges}
Adopting the notation introduced above, an isotropic shift line demands,
\beq\label{eq:iso2}
\frac{{^{63}\hat{K}}_{\perp}\left[\erww{\zeta_2}\right]-{^{63}\hat{K}}_{\perp}\left[\erww{\zeta_1}\right]}{{^{63}\hat{K}}_{\parallel}\left[\erww{\zeta_2}\right]-{^{63}\hat{K}}_{\parallel}\left[\erww{\zeta_1}\right]} \equiv \frac{\Delta_\zeta {^{63}\hat{K}}_\perp}{\Delta_\zeta {^{63}\hat{K}}_\parallel}\approx 1.
\eeq
We can always assume that this is true for $\zeta$ as an average over some local scale (see below), and the isotropic shift line is dominated by the isotropic shift component according to \eqref{eq:two1},
\beq
\Delta_\zeta {^{63}\hat{K}}_{\bot,\parallel} = {B} \cdot \Delta_\zeta \chi_2 \label{eq:iso2}.
\eeq
This means that this susceptibility is a function of $\zeta$, $\chi_2(\zeta)$. Then, a similar equation must hold for planar oxygen, as well, i.e. we expect,
\beq
\Delta_\zeta {^{17}K}_{\bot,\parallel} = F_{\bot,\parallel} \cdot \Delta_\zeta \chi_2 \label{eq:oxiso1}.
\eeq
We define $E_\alpha$ and $F_\alpha$ as the two hyperfine coefficients that relate the two spin component to the spin shift,
\begin{gather}
{^{17}{K}}_{\alpha} = {E}_{\alpha} \cdot \chi_1+{F}_{\alpha} \cdot \chi_2\label{eq:oxshift10}\\
{^{17}{K}}_{\alpha} = {E}_{\alpha} \cdot (a+c) +{F}_{\alpha} \cdot (b+c).\label{eq:oxshift11}
\end{gather}
Note that the orbital shift can be neglected for O \cite{Mangelschots1992}.

Early {$^{17}$O}~NMR experiments focussed on the question whether the planar O and Cu \emph{temperature dependent} shifts track each other. For \ybco \cite{Takigawa1989} and \ybcoE \cite{Bankay1994} this was found to be the case for all planar O shifts, but only $^{63,65}K_\perp (T)$ since the temperature independent $K_\parallel$ for \ybco was assumed to be orbital shift.

More elaborate research into the \emph{doping dependence} of the shifts appeared almost 20 years ago, by Haase et al. \cite{Haase2000,Haase2001,Haase2002}, but a deeper understanding could not be reached. In these publications an unexpected, direct relationship between the planar O magnetic shift and quadrupole splitting was reported. If one remembers that the planar O shifts in typical magnetic fields of \SI{\sim8}{Tesla}, have a similar size as the doping related changes of the quadrupole splitting, a correlation between both can lead to the reported peculiar effects that we review now. \par\medskip

The first effect concerns the average shift \cite{Haase2002} (it was noticed after the asymmetric lineshapes \cite{Haase2000} that we discuss as the second effect further below).

The $2I+1=5$ transitions of planar O ($I=5/2$), in leading order in the Zeeman interaction, can be written in frequency units (which are more useful for this purpose since the quadrupole splitting does not depend on field) in this simple form,
\beq\label{eq:ox1}
{^{17}f}_{\parallel,m}={^{17}f_\parallel}+m\cdot{^{17}\nu}_{\rm Q,c}\;,
\eeq
where $^{17}f_\parallel = {^{17}K}_\parallel  B_0$ represents the total shift in frequency units, $B_0$ is the external field; $\nu_{\rm Q, c}$ is the quadrupole splitting from \eqref{eq:oxfreq}, and $m = 0, \pm 1, \pm 2$ denotes the particular transition ($m=0$ is the central transition). Since only spectra for \cpara were recorded, we do not consider ${^{17}f}_\perp$. 

\begin{figure}
\centering
\includegraphics[width=0.45\textwidth ]{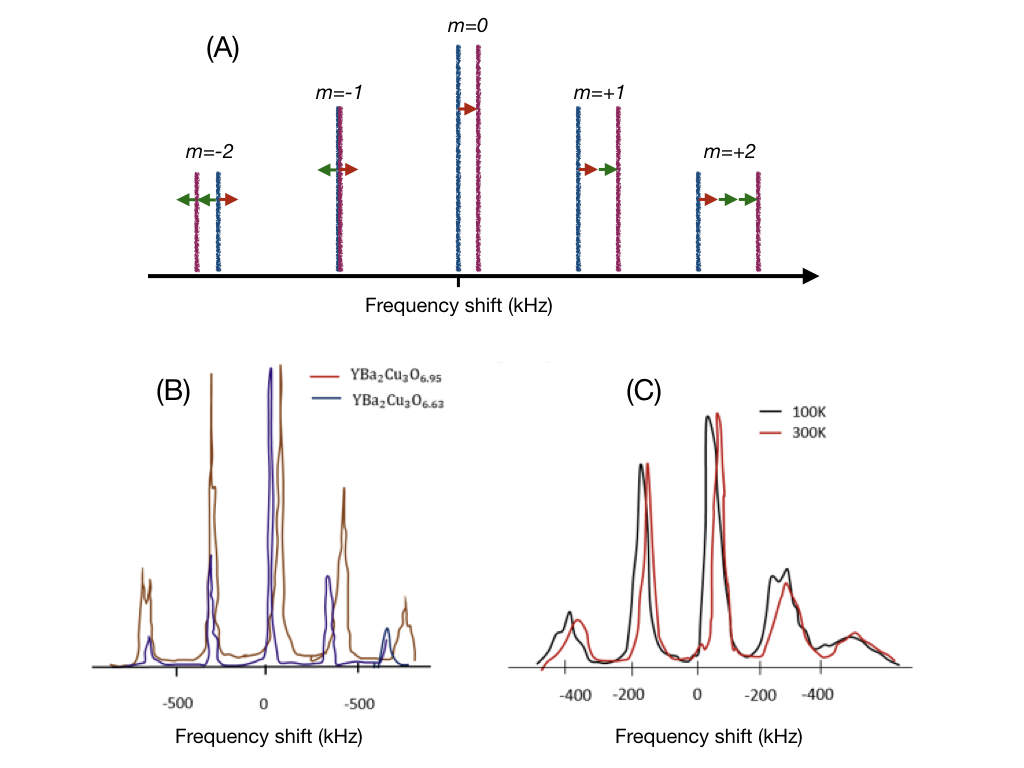}
\caption{ (A) Schematic of the high-temperature planar~{$^{17}$O} NMR spectra. The red arrows of equal lengths denote magnetic shift that is the same for all transitions. The blue arrows indicate the quadrupole splitting according to \eqref{eq:ox1}. In magnetic fields of $\SI{\sim8}{T}$ both contribution cancel for the left, first satellite ($m=-1$) in the cuprates. (B) Spectra for underdoped YBa$_2$Cu$_3$O$_{6.63}$ and optimally doped YBa$_2$Cu$_3$O$_{6.95}$ \cite{Haase2002x}. Note that the left satellite position does not change as a function of doping (the double peak pattern due to intra unit cell charge variation is explained in the text). (C) As a function of temperature, here shown for La$_{2-x}$Sr$_x$CuO$_4$ this is not the case, but the asymmetry is due to the same effect (data from \cite{Haase2001}).}\label{fig:fig5}
\end{figure}

It is important to note that the magnetic shift is the same for all transitions while the quadrupole splitting is not, as it depends on $m$ that also changes sign. This is illustrated in~Fig.~\ref{fig:fig5}(A).

It was observed, first for \lsco \cite{Haase2002}, that if one plots the total high-temperature planar O spectra for different doping levels $x$ (at least form $x=0.10$ to $x=0.20$) on top of each other, the $m=-1$ satellite does not change its position. This is a very striking experimental result, and it says,
\beq\label{eq:oxfix}
{^{17}f}_{\parallel,-1}(x)={^{17}f_\parallel}(x)-{^{17}\nu}_{\rm Q, c}(x) = const.
\eeq
With other words, over a large range of doping, the only change of shift at high-temperature is the same as the change in quadrupole frequency. However, lowering the temperature changes the constant in \eqref{eq:oxfix}, i.e. ${^{17}f}_{\parallel,-1}$. While the constant is independent of $\erww{\zeta}$ it depends on temperature, which is shown in Fig.~\ref{fig:fig5}(C). Available spectra for \ybco show the same behavior, cf. Fig.~\ref{fig:fig5}(B), for more evidence see further below.

Today, we know with \eqref{eq:x} how $x$ is related to $n_p$, i.e., it is the charge at planar O that changes linearly with $x=\erww{\zeta}$, which in turn leads to a proportional change of the splitting,
\beq\label{eq:R1}
\Delta {^{17}\nu}_{\rm Q, c} = R\cdot\Delta\erww{\zeta}
\eeq
according to \eqref{eq:oxfreq}, with
$$R = \frac{d {^{17}{\nu_{\rm Q,\parallel}}}}{d\zeta} = \SI{1.27}{MHz}\;\frac{d n_p}{d\zeta},$$
and we can estimate $d n_p/d\zeta$ from $\partial n_p/\partial \zeta =1/2$ and the slopes of $\Delta n_d/(2 \Delta n_p)$ in figure 5 of \cite{Jurkutat2014} (for \lsco we obtain $d n_p/d\zeta = 1/2.4$). 

This proves, cf.~\eqref{eq:oxfix}, that at high temperatures the planar oxygen shift ${^{17}K}$ changes linearly with $\erww{\zeta}$, as well,
\beq\label{eq:M0}
\Delta_\zeta {^{17}K}_{\parallel} = M \cdot \Delta \erww{\zeta},
\eeq
where $M$ is the proportionality factor. This is independent proof for a high-temperature doping dependence of the shifts as postulated based on the Cu data alone.

Based on the planar O data alone, we know that one or both of the susceptibilities in \eqref{eq:oxshift10} must be changing linearly with $\erww{\zeta}$. 
From the Cu NMR of systems that lie on an isotropic shift line we know that it is $\chi_2(\zeta)$ that causes this behavior, which leads us to the conclusion,
\beq\label{eq:M1}
\Delta_\zeta {^{17}K}_{\parallel} = F_\parallel \cdot\Delta_\zeta \chi_2,
\eeq
where $F_\parallel$ is the hyperfine constant from \eqref{eq:oxshift10}.
With \eqref{eq:oxfix}, \eqref{eq:R1} and \eqref{eq:M0} we have the relation,
\beq\label{eq:oxfix2}
{^{17}f}_{\parallel,-1}=f_0-\nu_0+(M-R)\erww{\zeta} = const.,
\eeq
with $M=R$ at high temperatures, and $f_0$ and $\nu_0$ being the magnetic shift and quadrupole splitting at the lowest doping level where \eqref{eq:oxfix2} holds, respectively. 

Before we continue the discussion, we point to the 2nd observation \cite{Haase2000}.

\begin{figure}[H]
\centering
\includegraphics[width=0.3\textwidth ]{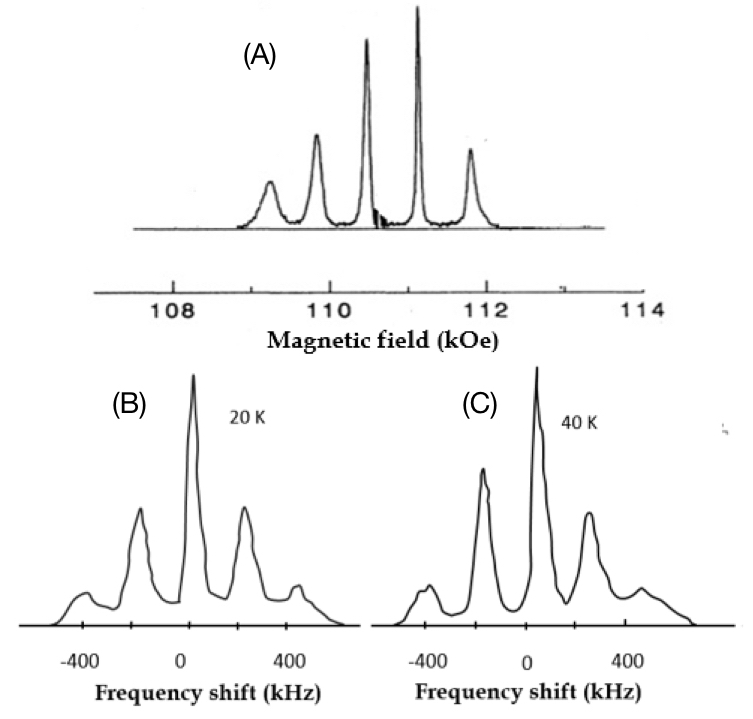}
\caption{(A) Field swept spectrum of the planar O in Tl$_2$Ba$_2$CuO$_y$ \cite{Kambe1993} shows the asymmetry (note that the ${m=-1}$ satellite appears to the right, due to a reversed frequency axis). (B, C) The asymmetry for La$_{1.85}$Sr$_{0.15}$CuO$_4$ begins to vanish just above the superconducting state due to the loss of correlation between spin shift and local charge \cite{Haase2001}.}\label{fig:fig6}
\end{figure}

This second peculiar effect concerns the $^{17}$O NMR lineshapes that are quite asymmetric, cf.~Figs.~\ref{fig:fig5} and \ref{fig:fig6}. It was observed, but not discussed already in the first publications on \ybco (Takigawa et al. \cite{Takigawa1989c} in their figure 1 the satellite resonances appear to have similar heights, but their amplification factors given underneath the spectra reveal that the $m = - 1$ satellite is much larger). For \ce{Tl2Ba2CuOy}, Kambe et al. \cite{Kambe1993} show similarly distorted spectra, cf.~Fig.~\ref{fig:fig6}. For \lsco the asymmetry was discovered and discussed in greater detail by Haase et al. \cite{Haase2000}. It was shown that this asymmetry can be understood by assuming a linear expansion of shift and splitting in some local parameter $h$ that is distributed across the \ce{CuO2} plane, in their notation \cite{Haase2000}, 
\begin{gather}
{^{17}{f}}_{\parallel}(h)=f_0 + M \cdot h \label{eq:z1}\\ 
{^{17}{\nu_{\rm Q,\parallel}(h)}}=\nu_0 + R \cdot h \label{eq:z2}
\end{gather}
where
\begin{equation}
M={d{^{17}{f}}_{\parallel}}/{dh}, \;\;\; R={d {^{17}{\nu_{\rm Q,\parallel}}}}/{dh},\label{eq:RandM}\\
\end{equation}
which for $h = \erww{\zeta}$ gives \eqref{eq:R1}, \eqref{eq:M1}, and eventually \eqref{eq:oxfix2}.

In the original paper it was also shown that one can use the distribution function of $h$ as obtained from the central transition lineshape (which is entirely magnetic), and predict the whole spectrum for optimally doped \lsco and \ybco with the ratio $R/M \approx 2$, not $R/M \approx 1$. 
If one sets $h = \zeta$, one has to note that a spatial variation of $\zeta$ modulates the planar O hole content $n_p$ that, in turn, changes the quadrupole splitting. While the latter is a very local property, this is not likely to be the case for the spin shift so that $R/M$ may depend on the wavelength of spatial oscillation of the charge, and $R/M$ is expected to be larger at the shortest wavelengths. We believe that the discrepancy between the linewidths and average splittings and shifts are to be expected. 

We note that while there have been only a few studies with respect to the overall independence of the $m=-1$ satellite, the effects behind \eqref{eq:oxfix2}, \eqref{eq:z1} and \eqref{eq:z2} describe the same phenomenon and thus support each other.

Another important early finding \cite{Haase2000} revealed that the magnetic linewidth associated with $M$ has a very different temperature dependence compared to that of the mean spin shift. In fact, we see in Fig.~\ref{fig:fig6} that the lines become symmetric only near entering the superconducting state (below \tc significant magnetic linewidth remains that is still in excess of what one expects from the fluxoid lattice \cite{Haase2001}). Similar observations were made for \hgryb \cite{Rybicki2012}, i.e., the magnetic linewidth associated with the charge variation does not follow the temperature dependence of the shift, but can decrease below \tc. 

We return to relative shift units and just write,
\beq\label{eq:oxfreq10}
{^{17}\hat{K}}_{\parallel} =f_0 + M' \erww{\zeta},
\eeq
($M' = M/B_0$), and if we take the difference between the central transition shift for two doping levels, we have, $${^{17}\hat{K}}_{\parallel}(\erww{\zeta_2})-{^{17}\hat{K}}_{\parallel}(\erww{\zeta_1}) =  M' (\erww{\zeta_2} - \erww{\zeta_1}),$$ which is independent of $f_0$. Indeed, for the available data we find that this difference does not change significantly above \tc. We conclude that $f_0$ must carry the temperature dependence at higher temperatures.\par\medskip

We repeat the conclusions to clarify the findings. First, we know that $(M'\cdot\erww{\zeta})$ is a doping dependent and temperature independent shift term (at high temperatures), and $f_0$ is a doping independent shift term that can carry a temperature dependence already at higher temperatures. Thus, there must be a spin susceptibility with $\Delta_\zeta \chi = M' \Delta \erww{\zeta}$ at high temperatures, and we know from the Cu shifts that it is $\chi_2$ since it can be responsible for isotropic shift lines, and thus couples to the Cu nuclear spin through $B$. Note, however, we cannot simply identify $(M'\cdot\erww{\zeta})$ with $\chi_2$, and $f_0$ with $\chi_1$. For example, with $\chi_2 = (b+c)$ the doping dependence could be carried by $b$, only, while $c$ could be temperature dependent already at higher temperatures. Similarly, $\chi_1= (a+c)$ could have a temperature and doping independent term $a$ while $c$ is temperature dependent already at higher temperatures. Alternatively, the sum $(b+c)$ could be doping dependent while $(a+c)$ is not \cite{Avramovska2019}.

We can be certain, however, that the planar O data demand two spin components.

\subsection{Discussion}
Since the planar Cu NMR shifts are very reliable and they have been investigated for many different cuprates, the new, overall phenomenology \cite{Haase2017} cemented the failure of the traditional view of the cuprates from NMR, which was summarized in the first part of the manuscript. A main feature in this new phenomenology is the appearance of an \emph{isotropic shift line} at high enough temperatures, i.e. both  Cu shifts change at the same rate as a function of doping, $\Delta_x {^{63}\hat{K}}_\perp/\Delta_x{^{63}\hat{K}}_\parallel \approx 1$. It was shown that orbital shift variations with doping cannot explain this finding, rather, an electronic spin with an isotropic coupling coefficient ($B$) is the only suitable candidate, in addition, it is expected on general grounds. However, such an explanation is in striking disagreement in particular with the old assumption for \ybco, that is ${^{63}\hat{K}}_\parallel = K_{L,\parallel}$, which was the basis of the traditional NMR scenario. 

\begin{figure}
\centering
\includegraphics[width=0.35\textwidth ]{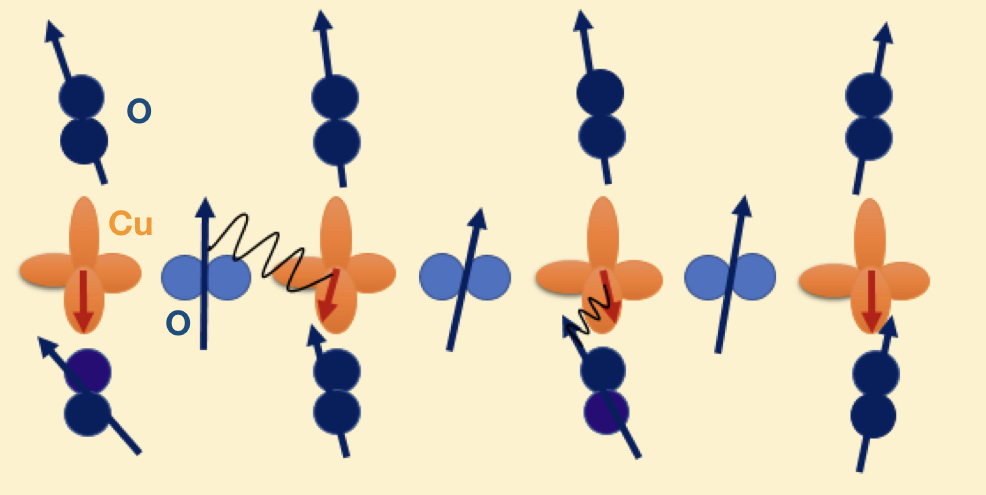}
\caption{Schematic of the intra unit cell charge variation of planar oxygen \cite{Reichardt2018} that occurs in addition to longer range charge density variations, in particular on oxygen. The total spin polarization as measured by a nucleus shows a variation, as well, and it is correlated with that of the charge, while perhaps over a larger scale compared to a unit cell.}\label{fig:fig7}
\end{figure}
Since the doping plays a special role in the description of the data, we summarized recent progress in the understanding of the charge distribution in the \ce{CuO2} plane. For example, we know that $x = \erww{\zeta}$, where $\zeta = n_d + 2n_p-1$ is the local charge per planar unit cell, and that $n_p$, only, sets the oxygen quadrupole splitting, cf. \eqref{eq:oxfreq}. 

Finally, we investigated the validity of such a two-component approach by exploring the $^{17}$O NMR shifts and quadrupole splittings. Here, we find that the peculiar, yet ubiquitous behavior of planar $^{17}$O NMR, which was reported some time ago \cite{Haase2000,Haase2002}, but not understood, actually demands a two-component scenario much of the kind suggested recently \cite{Haase2017,Avramovska2019}. 

In those studies \cite{Haase2000,Haase2002}, it was observed that the quadrupole splittings and high temperature shifts are intimately linked for planar O, documented by two related observations: a doping-independent $m=-1$ satellite, and special asymmetric total lineshapes. Both are a manifestation of a correlation of $n_p$ with $\chi_2$, cf.~Figs.~\ref{fig:fig5} and \ref{fig:fig6}. 

Thus, just based on planar O data we concluded on a doping dependent susceptibility $\chi_2(\zeta)$ that increases linearly with $\zeta$ and that is temperature independent at higher temperatures, very similar to what was concluded from the Cu data \cite{Haase2017,Avramovska2019}, earlier. However, the full temperature dependence of $\chi_2(\zeta)$ is not well known, apart from being temperature independent at higher temperatures and its vanishing at the lowest temperatures. 
The more elaborate data set for the \hgryb family clearly showed this component \cite{Rybicki2015}, but we believe that the low temperature value is not entirely of orbital origin, which changes some conclusions made in that publication \cite{Rybicki2015} (the possibility was, however, considered \cite{Rybicki2015}). Extensive experiments for \ybco were reported \cite{Machi2000} that showed that the planar O NMR shift anisotropy is rather temperature independent even below \tc, but does not track the isotropic shift, in support of our findings.

From the temperature dependences of the shifts alone it is difficult to determine more details, except that the planar $^{17}$O NMR shift can become negative for various systems \cite{Takigawa1991,Takigawa1989,Yoshinari1990}, as expected for a negative $\chi_1$  \cite{Haase2017,Avramovska2019}. However, the measurements of the shifts below \tc are very sparse, in particular down towards the lowest temperatures. 

The asymmetric planar O spectra are based on the same phenomenon. Here a spatial variation of $n_p$, and with it that of $\chi_2(n_p)$ leads to the special asymmetric lineshapes.
This also explains the quite general observation that magnetic and quadrupolar linewidths appear together in  cuprate NMR. A spatial variation of $n_p$ also affects the planar Cu NMR linewidth, cf.~\eqref{eq:cufreq}, but the same equation shows that a variation of $n_d$ is more effective. The materials with the most narrow NMR lines are the stoichiometric systems, but these exhibit split oxygen satellite resonances that have been proven to be caused by intra unit cell charge density variations \cite{Reichardt2018}. 
The difference in quadrupole splittings point to a total O hole variation of $\sim 1\%$. The difference in the magnetic shifts between these two oxygen sites within the unit cell is less than what one expects from the mere difference in charge \cite{Haase1999x}, which is to be expected due to the small length scale. It appears that this intra unit cell charge variation is the source for the instability towards charge modulations in the \ce{CuO2} plane \cite{Jurkutat2019b}.
A possible sketch of the charge and spin density variation is depicted in Fig.~\ref{fig:fig7}. 

Again, NMR of planar Cu and O demands a doping dependent shift term that is quite unusual. In addition, the second shift term that remains  even below \tc is very unusual, as well, and it points to a magnetic ground state that involves Cu $3d(x^2-y^2)$ spin. Since the nuclear relaxation freezes out universally in the cuprates \cite{Avramovska2019,Jurkutat2019}, APH argued that relaxation originates from $b$ and that the doping dependence is caused by the negative coupling $c$ with $(a+c)$ being independent on doping. However, we believe it is also possible that doping changes predominantly $b$ at higher temperature, and that $a$ is responsible for a doping independent universal Cu relaxation. In such a scenario a change in $b$ would explain the observed change in the relaxation anisotropy. Clearly, one has to demand that the fluctuations $\erww{a(t)a(t+\tau)}$ need to freeze out below \tc, together with those of $\erww{b(t)b(t+\tau)}$. 
The exchange coupling $c$ turns on the pseudogap effect and is perhaps involved the paring, as well.

Finally, we see no reason to challenge the explanation of NMR in terms of two spin components, given the fact that components with $A_\alpha$ and $B$ are to be expected in the material from solid chemistry arguments. Furthermore, the suggested scenario also solves the orbital shift conundrum. To what extent loop currents \cite{Varma1997} can fit into such unusual shifts is not clear to us. One should also note that measurements of the uniform spin susceptibility were not aware of a negative spin component, as far as we know, and they might contribute to a better understanding of our scenario.

\subsection{Conclusions}
The planar $^{17}$O NMR shifts demand a two spin component scenario, since the shifts are given by ${^{17}{K}}_{\parallel} =f_0 + M' \erww{\zeta}$, where $M'$ is temperature independent at high temperatures, while $f_0$ is responsible for the high-temperature (pseudogap) spin shift, which was discovered many years ago \cite{Haase2003}. With the recent advance in understanding the charge sharing in the cuprates \cite{Jurkutat2014} it could be concluded that changes in $\erww{\zeta}$ are given by doping related changes of the planar O charge in the $2p_\sigma$ orbital. 

These findings have great similarity with the recently established two-component scenario that was based predominantly on planar Cu NMR shifts \cite{Haase2017,Avramovska2019}, where two spin components $a, b$ and a coupling term $c$ describe the spin shifts ${^{63}{K}}_{\parallel,\perp}=A_{\parallel,\perp} (a+c)+ B (b+c)$. It was found that $\Delta_\zeta (b+c)$ varies with doping as it can lead to isotropic shift changes as a function of doping. One then expects that the planar O shifts obey a similar equation, ${^{17}{K}}_{\parallel,\perp}=E_{\parallel,\perp} (a+c) + F_{\parallel,\perp} (b+c)$. It is possible that $(b+c)$ is related to $(M'\cdot \erww{\zeta})$, the doping dependence could also be carried by $b$ alone. More precise planar O shift data, in particular also at very low temperature could help in a definite assignment.

We showed that a spatial variation of the local charge and thus of the oxygen hole content $n_p$, together with the doping dependent $\chi_2(n_p)$, lead to correlated magnetic and quadrupolar linewidths, common in the cuprates. This spatial variation of $n_p$ can be affected by crystal chemistry, but it cannot be reduced below the intrinsic, intra unit cell planar O charge variation \cite{Haase2003,Reichardt2018}. 

The unconventional shift of planar Cu and O have the same origin and must be relevant to the understanding of the cuprates above and below \tc.

\subsection*{Acknowledgements}
We acknowledge support from Leipzig University.

\subsubsection*{Author contributions}
J.H. introduced the main concepts and had the main leadership; all authors were involved in data analysis and discussion equally, as well as in preparing the manuscript.

\vspace{0.5cm}

\bibliography{JHPRL.bib}   

\printindex
\end{document}